\documentclass[12pt]{article}
\usepackage{amsmath, amsfonts}
\usepackage[dvips]{graphics}
\newcommand{\bee}{\begin{eqnarray*}}
\newcommand{\ene}{\end{eqnarray*}}
\newcommand{\beeq}{\begin{equation}}
\newcommand{\eneq}{\end{equation}}
\newtheorem{lem}{Lemma}[section]
\newcommand{\bel}{\begin{lem}}
\newcommand{\enl}{\end{lem}}
\newtheorem{defi}{Definition}[section]
\newcommand{\bef}{\begin{defi}}
\newcommand{\enf}{\end{defi}}
\newtheorem{exap}{Example}[section]
\newcommand{\beex}{\begin{exap}}
\newcommand{\enex}{\end{exap}}
\newtheorem{theo}{Theorem}[section]
\newcommand{\beth}{\begin{theo}}
\newcommand{\enth}{\end{theo}}
\newtheorem{prop}{Proposition}[section]
\newcommand{\bep}{\begin{prop}}
\newcommand{\enp}{\end{prop}}
\newtheorem{cor}{Corollary}[section]
\newcommand{\bec}{\begin{cor}}
\newcommand{\enc}{\end{cor}}
\newtheorem{rem}{Remark}[section]
\newcommand{\ber}{\begin{rem}}
\newcommand{\enr}{\end{rem}}

\usepackage{amsmath}
\usepackage{graphicx}
\usepackage{latexsym}
\begin{document}
\title{ 
%\begin{center}
%{
%PRIVATE COMMUNICATION/NOT FOR CIRCULATION\\
%{\bf 
%BAYES RISK\\ AND \\ THE PRICE OF A EUROPEAN OPTION
%THE  BAYES 
%PRICE\\ OF \\ 
%EUROPEAN 
%CALL OPTION, \\  
%ARBITRAGE,\\ 
%THE BLACK-SCHOLES-MERTON PRICE\\
%THE PRICE  OF 
%EUROPEAN
%A CALL OPTION,\\
OPTION PRICING,\\
BAYES RISK\\
AND \\ APPLICATIONS
%(STATISTICAL FINANCE I)
%THE DISTRIBUTION OF 
%(STATISTICAL FINANCE I: MODELING) 
%}
}
%\\
%by\\
\author{Yannis  G. Yatracos\\
%Cyprus U. of Technology and National U. of Singapore
Cyprus U. of Technology 
}
%\end{center}
\maketitle
\date{}
%Department of Statistics and Applied Probability,
%The National University of Singapore, 6 Science
%Drive 2, Singapore 117546, {\em e-mail:} yatracos@stat.nus.edu.sg
%\bigskip
%%%%%%\vspace{3in}
%{\em Running head:} Modeling stock prices with Le Cam's experiments\\
%\vspace{.2in}
%{\em AMS 1991 subject classifications:} Primary ; Secondary

%\vspace{.1in}
%Address: Department of Statistics and Applied Probability,
%The National University of Singapore, 6 Science
%Drive 2, Singapore 117546 \\ 
%Running Head: Option pricing with Le Cam' s  experiments\\
%\mbox{ } \\
%{\it e-mail:} stayy@nus.edu.sg, Yannis.Yatracos@cut.ac.cy
{\it e-mail:} yannis.yatracos@cut.ac.cy 

%JEL classifications: C11, C12, C44, G13

%AMS classifications: 62, 60

\pagebreak 
 
\begin{center}
{\bf Summary}
\end{center}

%The  Bayes ({\it B}-) price  of the European call option 
%at time $t_0$  
%is introduced. 
%Within a class 
%${\cal C}$ of potential option prices, 
%{\it B}-price is that closest to the stock's price  at $t_0$
%and ${\cal C}$'s largest. 
%Calculated with the equivalent martingale probability,
%{\it B}-price does not allow arbitrage and is shown to be the ``fair'' price.
%The results provide new interpretations for the ``fair'' price.
A statistical decision problem is hidden 
%, surprisingly, 
  in the core of option pricing.
% is a statistical decision problem.
% by providing a
A simple form  for the
%`fair''
 price $C$ of a European call option  is  obtained
 via  the minimum Bayes risk, $R_B,$ of a 2-parameter  estimation
problem,
% for {\it 0-1} loss, 
%and a given prior
 thus justifying calling $C$ 
%also
Bayes (B-)price. 
% The trader who sells a covered option can use $R_B$ as  index of its accounting leverage at the transaction time 
%and prefers ``small''  $R_B$-values.
  The result provides new insight in option pricing, among others 
 obtaining $C$ for some stock-price models  using the underlying probability
%without using
instead of
the risk neutral probability  and giving $R_B$ an  economic interpretation.
%making it accessible to a larger audience than the finance specialists.
% and
% suggest discounting stock prices with
%expectations' ratios. 
%B-prices are obtained
When logarithmic stock 
prices follow Brownian motion, discrete normal mixture
% For logarithmim stock prices that follow   
and hyperbolic L\'{e}vy motion the obtained B-prices are ``fair" prices.
%  B-price  complements the prices obtained using the Gerber and Shiu (1994) pioneering approach.
%For the latter it is seen that,
%for short maturity period $B$-price is the ``fair'' price and is
%smaller than the Black-Scholes-Merton price.
%The results indicate
% also  a 
A new expression for the price of American call option is also obtained and  
statistical modeling of $R_B$ can be used when pricing
% both 
European and American call options.
%When the market is not complete, a price of  European call option is suggested 
%that does not allow arbitrage based on the beliefs of the option's buyers.
 
\vspace{1in}


\pagebreak

%\begin{center}
%\section{\bf Bayes risk for 0-1 loss and the price of a European option} 
%\end{center} 
 \section{Introduction}

\quad So far, the core of  the option pricing problem is a topic studied mainly  in  Finance and Mathematical Finance.
This work shows, surprisingly, that option pricing can be seen as a statistical decision problem  with some useful implications. 
%A brief introduction of
%the option pricing problem follows for statisticians with minimal exposure in it.

 By purchasing 
%a European call option  for a share
at  time $t_0$ for premium $C$
a share's European call option,
%In a financial transaction 
the buyer has in the future, at time $T (>t_0),$ the option to buy the share at
predetermined fixed value $X.$ 
The ``fair'' price $C$ is obtained, under some assumptions,  by ``replicating the call,''  i.e. by creating a portfolio 
that matches the call's
payoff at $T$ (Black and Scholes, 1973,  Merton, 1973).  This procedure guarantees $C$ does not allow arbitrage, 
i.e. that  the call option's buyer  cannot  make profit with probability 1.
However, ``the procedure may be tedious and computationally demanding'' (Sundaram, 1997, p. 85) .
Alternatively,
% The method of risk neutral (or equivalent martingale) probability provides an alternative possibility: 
%It was shown that 
$C$ is obtained by discounting at $t_0$ the expected value of the call's payoff at 
%maturity
$T$  under  the risk neutral (or equivalent martingale)  probability that is not always easy to determine.
%identified from the modeling assumptions of the share's price. 
The interested reader may refer to Sundaram (1997)
for a rather informal 
and accessible introduction to the use and determination of  risk neutral probability.

More complex financial instruments have been introduced and priced in the sequel, 
as for example the American call option, where the option's  holder may exercise the right to buy the share at price $X$ 
any time $t \in (t_0,T];$ see Hull (1993) for other types of options.   
So far,  for the European, the  American and other call options, no statistical problems have been determined 
whose solutions provide the corresponding ``fair'' price.

%A decision theoretic approach to European
%option pricing leads to 
Using the equivalent martingale probability approach, a simple, new expression for the price $C$ of a European call option is 
obtained herein,
which involves the minimum Bayes risk, $R_B,$  of a 2-parameter statistical estimation problem and some known quantities.
In this way
%, ``fair'' 
 option pricing can be seen as the solution of a statistical problem. In fact,  $C$ increases when the corresponding statistical estimation problem becomes simpler.
%IS presented herein. 
The result suggests discounting stock prices with expectations' ratios.
$C$ is calculated via 
%Bayes risk 
$R_B$ for various stock price models, circumventing in some cases the search
for the equivalent martingale probability and complementing the approach 
by Gerber and Shiu (1994) when the martingale probability is not unique. 
%Bayes risk
For the trader selling the call option,
% at $t_0,$
 $R_B$ is a lower bound on  the ratio of its liability and its  expected assets at $t_0$,  called  ``accounting leverage".
% for the trader selling
%the call option
% and is also useful in
$R_B$ can be used when calculating  the price 
%calculation 
of an American option. 
%Motivation is also provided for
%subsequent modeling of the stock's logarithmic price returns via Le Cam's 
%statistical experiments and for the pricing of
%the stock's European call option (Yatracos, 2011).
%The Bayesian approach used in this work and the obtained results are not 
%related with  the Bayesian calibration framework used to mark the stock price 
%model to market, obtaining posterior distributions of model parameters for 
%a given prior distribution (see, for example, Gupta and Reisinger, 2011).    

The Bayesian approach in this work and the results are not related
 with  Bayesian calibration  used to mark the stock price model to
market obtaining posterior distributions of model parameters for a given prior
(see, for example, Gupta and Reisinger, 2011).
%The results motivate the use of Le Cam's statistical experiments in modeling stock price returns and in pricing the stock's European option %(Yatracos, 2011).
For a different  statistical approach to theoretical and practical issues in stock trading, including option pricing, the interested reader may consult Franke, J. et al. (2010).

In section 2, a quantitative description  of the results is presented. In section 3, the new expression for $C$  is obtained via $R_B.$ In the applications in sections 4 and 5, $C$  is obtained via $R_B$  when the logarithmic stock prices
follow Brownian motion, discrete normal mixture and hyperbolic  L\'{e}vy motion.
Proofs 
%not in section 
are in the Appendix and Figure 1  appears after the references.

\section{A quantitative description of the results}

\quad Let $S_t$ be the price of the stock at time $t, \ 0< t_0\le t \le T, 
\ S_{t_0}=s_{t_0},$  let
$(\Omega, {\cal F}, {P})$ be the underlying probability space and
let $P^*$ be the (assumed for now) unique martingale probability equivalent to $P.$ 
%$ES_{t_0}=s_{t_0}.$
The buyer of a European call option at $t_0$   has the right  to buy one share 
at  time 
%$T(>t_0)$
$T$
with  
``strike'' 
 price $X$  by paying  ``fair'' price $C,$ i.e. the discounted, $P^*$-expected
cost at maturity.
An
% general 
expression for $C$ is obtained that involves
cumulative distributions $F_1^*$ and $F_0^*$ defined 
%in  (\ref{eq:h0})  and (\ref{eq:F11cdf}) 
%obtained 
via $P^*; \ F_0^*$ is the cumulative distribution of 
$\ln \frac{S_T}{ES_T}$
% obtained 
under $P^*$ and $F_1^*$ is an 
equivalent probability determined in
%see (\ref{eq:h0})  and 
(\ref{eq:F11cdf}).
Distributions $F_1^*$ and $F_0^*$ constitute the parameter space of the statistical decision
 problem that determines the option's price.
It is shown that 
\begin{equation}
C=s_{t_0}-R_B(s_{t_0}+Xe^{-r(T-t_0)}), 
\label{eq:THEMAINRESULT1}
\end{equation}
with $R_B$
%is multiple of 
the minimum Bayes
risk 
%$R_B$ 
for the estimation 
%problem of 
%cumulative distributions 
of $F_1^*$ and $F_0^*$ with $0-1$ 
loss, 
%anda prior, 
$0<R_B<1, \ r=\ln (1+i), \ i$ fixed interest.
From (\ref{eq:THEMAINRESULT1}) it follows that when the difficulty of 
%the estimation problem increases, $R_B$ increases and the  
%$C$-value  decreases.
the estimation problem decreases by increasing the Hellinger distance between $f_0^*$ and $f_1^*, \ R_B$ decreases and the  
$C$-value  increases. For example, under the 
Black-Scholes-Merton ({\it B-S-M}) assumptions for the stock price process,
when the volatility increases
the Hellinger distance between $f_0^*$ and $f_1^*$ increases, therefore
the difficulty of the estimation problem decreases and hence the $C$-value 
increases.
It also follows that $C$  can be obtained from a simple game with
loss,  profit  and
% and it is the ``fair'' price of
%a game with loss and profit and 
%%with 
%corresponding 
respective probabilities determined by 
$R_B.$ 

%Accounting leverage a
At $t_0,$ accounting leverage for the trader writing
% a covered 
the call option by taking a loan
%and having expected asset $s_{t_0}+ 
can be measured with the ratio
\begin{equation}
\label{eq:leverage}
\frac{s_{t_0}-C}{s_{t_0}+Xe^{-r(T-t_0)}P(S_T>X)}\ge R_B=\frac{s_{t_0}-C}{s_{t_0}+Xe^{-r(T-t_0)}}.
\end{equation}
The numerator in the left side of
 (\ref{eq:leverage}) is the trader's liability and the denominator its total expected assets, both at $t_0.$ 
Since $P(S_T>X)$ is unknown, the trader or the bank providing the amount $s_{t_0}-C$ may not allow the transaction when  $R_B$ is 
``high'' .   

The results justify naming $C$ 
Bayes ({\it B}-) price denoted by $\tilde C_{B,t_0}(P^*)=C$
and
%one can compute {\it B}-price $\tilde C_{B,t_0}(P)$ under (any) $P; 
%\ \tilde C_{B,t_0}(P^*)=C$
%The minimum Bayes risk $R_B$ is a lower bound on the leverage  of the 
%call's writer 
%who intends to buy a share at $t_0;$ the leverage 
%is the ratio of its liability 
%$s_{t_0}-C$ and assets at $t_0.$
%The results 
indicate a new expression for the price of an American option in
Karatzas (1988, p. 50, equation 5.10 with 
parameter values those in Example 4.4, p. 44).
%$d=1, \ g(t)=0, \ r(u)=r, 
%\f_t=(P_1(t)-c)^+$). 
Advantages of (\ref{eq:THEMAINRESULT1}) include 
its simplicity and the possibility of 
modeling  $R_B$ in order to obtain $C$ or its approximation for various 
choices of $F_0^*$ and $F_1^*.$
%When there are several martingale probabilities equivalent to $P$
%one can obtain several ``fair'' Bayes prices and associated Bayes risks and
%the largest price does not allow arbitrage; see Remark \ref{r:manyEMM}.

Motivated by these findings and the definition of $F_0^*,$
we discount stock prices with  expectation ratios
and it is shown 
under {\it B-S-M} assumptions
that  {\it B}-price $\tilde C_{B,t_0}(P)$
is 
{\it B-S-M} price $C$  obtained under $P^*.$
This is not surprising since
the  expectations ratio discounted prices 
%(EDP) 
$\{\frac{E_PS_{t_0}}{E_PS_t}S_t=\frac{s_{t_0}}{E_PS_t}S_t, t \ge t_0\}$ 
are martingale under $P.$ 
%\ E$ denotes expected
%value with respect to $P.$
Note also that  for the equivalent martingale probability $P^*$ of {\em any} $P,$
$\frac{s_{t_0}}{E_{P^*}S_t}=e^{-r(t-t_0)}$ i.e. the usual discounting factor.
%with $r=\ln (1+i)$ and $i$ fixed interest.

Since  discounted price $s_{t_0}S_T/E_PS_T$ is used when calculating 
{\it B}-price,
sufficient conditions are  provided
for the mean-adjusted process $\{S_t/E_PS_t, \ t>0\}$ of geometric prices
to be martingale under $P.$ These conditions hold for all $t$
when $\ln (S_t/E_PS_t)$
is Brownian or hyperbolic L\'{e}vy motion, but for discrete normal mixture
$\{S_t/E_PS_t, \ t>0\}$ 
is ``nearly'' martingale for small $t,$ which is sufficient to obtain the option's price.
{\it B}-price for the latter, with small $T$-values, indicates overpricing
when using instead {\it B-S-M} price for normal distribution with 
the same mean and variance. {\it B}-price for L\'{e}vy motion is a ``fair'' price complementing the 
prices obtained by Eberlein and Keller (1995)
using the Esscher transform (Gerber and Shiu, 1994).

%The Bayesian approach in this work and the results are not related
% with  Bayesian calibration  used to mark the stock price model to
%market, obtaining posterior distributions of model parameters for a given prior
%(see, for example, Gupta and Reisinger, 2011).
%The results motivate the use of Le Cam's statistical experiments in modeling stock price returns and in pricing the stock's European option %(Yatracos, 2011).
%For a different  statistical approach to theoretical and practical issues in stock trading, including option pricing, the interested reader may %consult Franke, J. et al. (2010).
%Proofs 
%are in the Appendix and Figure 1 after the references.

%\section{The fair price of  European call option and Bayes risk and price}
\section{The  price of a  European call and Bayes risk}

\quad {\it SET-UP ({\cal A}): $S_t$ is the stock's price at time $t$
on the probability 
space $(\Omega, {\cal F},P),
 \ 0<t_0\le t \le T, S_{t_0}=s_{t_0}; \ X$ is the strike price at maturity $T;
\ P^*$ is the unique equivalent martingale probability to $P;$
%at $t_0,$ $X$ is the ``strike'' price of the European call option at 
%maturity $T;$
$ r=\ln(1+i), \ i$ is fixed interest;  probabilities 
$\pi_1, \ \pi_0$ are
$\pi_1=\frac{s_{t_0}}{s_{t_0}+Xe^{-r(T-t_0)}}=1-\pi_0;$
expectation $EU$ is obtained
under $P^*.$ The ``fair'' price $C$ of a European call option is the
discounted expected value of the call's payoff at maturity under $P^*:$
\begin{equation}
C=e^{-r(T-t_0)}E(S_T-X)I(S_T>X);
\label{eq:FAIRPRICESETUPA}
\end{equation}
$I$ denotes indicator function.
}

Denote by   $F_0^*$  the c.d.f.  of 
$Y=\ln \frac{S_T}{ES_T}$ under $P^*,$
\begin{equation}
%%% P[\ln \frac{S_T}{a(t_0,T)} \le y]=F_0(y), \  F'_0(y)=f_0(y).
P^*(Y \le y)=P^*[\ln \frac{S_T}{ES_T} \le y]=F^*_0(y), \  F^{*'}_0(y)=f^*_0(y), 
\ -\infty<y<+\infty. 
\label{eq:h0}
\end{equation}
Observe that 
\[
%\begin{equation}
%\label{eq:F1cdf}
1=E\frac{S_T}{ES_T}=Ee^{\ln \frac{S_T}{ES_T}}=Ee^Y=\int_{-\infty}^{+\infty}
e^yf^*_0(y)dy,
%\end{equation}
\]
thus for 
\begin{equation}
\label{eq:F11cdf}
f_1^*(y)=e^yf_0^*(y), \ F_1^{*'}(y)=f_1^*(y), \ -\infty<y<+\infty,
\end{equation}
it follows that $f_1^*$  is density with cumulative
distribution function $F_1^*$ and the mean value under $F_0^*$ is smaller than that under $F_1^*.$

%%%PART 2 BEGGINING-TO KEEP

%RELATING $f_0$ with $h_0$ and showing $f_1$ is a density

%Let $Y=\ln \frac{S_T}{ES_T}.$ Then, the density of $Y,$
%$$f_0(y)=h_0(e^y ES_T)e^yES_T,$$
%and if $f_1(y)=e^yf_0(y)$ then
%$$\int_{-\infty}^{\infty} f_1(y)dy=\int_{-\infty}^{\infty}  e^y f_0(y)dy=
%\int_{-\infty}^{\infty} e^y h_0(e^y ES_T)e^yES_T dy$$
%and with the transformation $s=e^y ES_T,$
%$$=\int_{-\infty}^{\infty} e^y h_0(s)s dy=\int_{-\infty}^{\infty} \frac{s}{ES_T}h_0(s)ds=1.$$

%%%PART 2 END

%It is seen below that the ``fair'' price (under 
%$P^*$) is determined by the minimum Bayes risk for {\it 0-1} loss
%of the estimation problem with parameters $F_0^*$ and $F_1^*$ and
%prior probabilities, respectively, $\pi_0$ and $\pi_1.$

\bep \label{p:P*fairisBayes}
Under ({\cal A}), (\ref{eq:h0}) and (\ref{eq:F11cdf}),
for the ``fair'' price $C$ of
the European call option at $t_0$ it holds\\
a)
\begin{equation}
\label{eq:CRB}
\frac{s_{t_0}-C}{s_{t_0}+Xe^{-r(T-t_0)}}=\pi_1F^*_1(-D)+\pi_0 [1-F_0^*(-D)]
=\inf_{d>0} [\pi_1F^*_1 (\ln \frac{d}{ES_T})
+\pi_0 [1-F_0^*(\ln \frac{d}{ES_T})]],
\end{equation}
with $-D=\ln(X/s_{t_0})-r(T-t_0).$
%with maturity $T$ and strike price $X$ 
The right side of (\ref{eq:CRB}) is the minimum
Bayes risk $R_B$   
%(\ref{eq:Bayes}) 
for the hypotheses $F_0^*$ and $F_1^*$ under  {\it 0-1} loss
%and prior 
with probabilities, respectively, $\pi_0$ and $\pi_1.$ 
%(\ref{eq:prior01}).\\
The value for which the posterior densities of $F^*_0$ and $F^*_1$ are equal
determines $R_B.$\\
b) From a) it follows that
%For the  Bayes price $C=\tilde C_{B,t_0}(P^*)$ of the call option it holds
\begin{equation}
C=(1-R_B)s_{t_0}-R_B X e^{-r(T-t_0)}=
s_{t_0}-R_B(s_{t_0}+X e^{-r(T-t_0)})
\label{eq:Bayesprice}
\end{equation}
\begin{equation}
=s_{t_0}[1-F^*_1(-D)]-Xe^{-r(T-t_0)}[1-F^*_0(-D)].
\label{eq:Bayes2}
\end{equation}
%with
%$R_B=\pi_1F^*_1(-D)+\pi_0[1-F*_0(-D)], \ -D=\ln(X/s_{t_0})-r(T-t_0).$\\
Thus, $C$ can be called Bayes (B-) price, denoted also by 
$\tilde C_{B,t_0}(P^*).$
\enp

We revisit Schachermayer's (2008)
 ``toy'' example for an illustration.
%where the supremum of {\it B}-prices over all possible
% ``beliefs''-probabilities is the ``fair'' price.

\beex \label{ex:toy}
In
Schachermayer's (2008)  ``toy'' example, at time $t_0=0$
the stock has price $s_0=1 \ \mbox{USD} $
 and under martingale probability
$P^*,$
%with $P^*(\{g\})=1/3=1-P^*(\{b\})$
%for which  $E_QS_1=1$ with $S_0=s_0=1;$
\[ S_1= \left \{\begin{array}{ll}
2 \ \mbox{USD}& \mbox{with prob.  $P^*(S_1=2)=1/3,$} \\
.5 \ \mbox{USD}  & \mbox{with prob.  $P^*(S_1=.5)=2/3,$}
\end{array}
\right.\]
and $E_{P^*}S_1=1 \ \mbox{USD}.$

For strike price $X=1 \ \mbox{USD},$ the
%``fair''
price of the European option with maturity
$T=1$ and fixed interest $i=0$ is
$$E_{P^*}(S_1-1)_+=1/3 \ \mbox{USD}.$$
Let $d$ be a  generic exercise barrier like that used for Bayes risk in
%the right side of
(\ref{eq:CRB}).
%Consider a  generic exercise barrier
%$d.$
To obtain the {\it B}-price for $P^*$ maximize
over $d$
\begin{equation}
s_0E_{P^*}\frac{S_1}{E_{P^*}S_1}I(S_1 \ge d)-X P^*(S_1\ge d)
\label{eq:toyBayes1}
\end{equation}
$$
=E_{P^*}S_1I(S_1 \ge d)-P^*(S_1\ge d)
= \left \{\begin{array}{ll}
1-1=0 \ \mbox{USD}, & \mbox{for  $0\le d \le .5 $} \\
2 \cdot \frac{1}{3} -1 \cdot \frac{1}{3}=\frac{1}{3} \  \mbox{USD} & \mbox{for $.5<d\le 2.$}
\end{array}
\right.
$$
Thus,  {\it B}-price is the ``fair'' price  $\frac{1}{3} \ \mbox{USD}.$

\bec
\label{c:locscaprice}
In addition to the assumptions used in Proposition \ref{p:P*fairisBayes},
assume  that $F_0^*$ and $F_1^*$  are 
location-scale cumulative distribution functions,
 i.e.
\begin{equation}
F_0^*(y)=G_0(\frac{y-\theta_0}{\sigma_0}), \  F_1^*(y)=G_1(\frac{y-\theta_1}{\sigma_
1}),
\label{eq:locsca}
\end{equation}
$\theta_i \in R, \ \sigma_i >0, \ i=0,1,$ and that $G_i(x)=1-G_i(-x), \ x \in R,
 \  i=0, 1.$
%and $G_1$ are symmetric around zero.
Then,  the Bayes price of the call option is
\begin{equation}
C=\tilde C_{B,t_0}(P^*)=s_{t_0}G_1(\frac{D+\theta_1}{\sigma_1})
-Xe^{-r(T-t_0)}
G_0(\frac{D+\theta_0}{\sigma_0});
\label{Bayespricelocscale}
\end{equation}
where
\begin{equation} 
D=\ln (s_{t_0}/X)+r(T-t_0).
\label{eq:Bayesbarrier}
\end{equation}
\enc

\ber
\label{r:fromP*toP}
Under {\it B-S-M} assumptions, (\ref{Bayespricelocscale}) is 
the {\it B-S-M} price; see (\ref{eq:bsmprice}).
%When discounting $S_T$ with $\frac{s_{t_0}}{E_PS_T}, \ X$ with 
%$e^{-r(T-t_0)}$ and the price of the option is calculated
%under $P,$  
%Proposition 2.1 and Corollary 2.1 hold with $P^*, \ F_1^*, \ F_0^*$ 
%replaced by $P$ and 
%$F_1, F_0$ defined as in (\ref{eq:h0}) and  (\ref{eq:F11cdf}),
%but the obtained price is not necessarily the fair price.
\enr

\bec
\label{c:simplegame}
Under the assumptions in Proposition \ref{p:P*fairisBayes},
let G be the game that results in loss $-(s_{t_0}-\tilde C)$ 
with probability $1-R_B$ and profit
$\tilde C+Xe^{-r(T-t_0)}$
with probability $R_B$ (the Bayes risk).  
The value $\tilde C=\tilde C_{B,t_0}(P^*)$ makes $G$ ``fair''.
\enc

\ber
\label{r:American}
The results in Proposition \ref{p:P*fairisBayes} suggest for the American
call option with strike $X$ that can be exercised in $(t_0,T]$ the price
$\tilde C_{B,A}=s_{t_0}-\inf_{t \in (t_0,T]}
\{(s_{t_0}+Xe^{-r(t-t_0)})R_{B,t}\}; 
\ R_{B,t}$
is the 
Bayes risk under $P^*$
of the European option with maturity $t, \ t_0<t\le T.$ The obtained
$\tilde C_{B,A}$ is a different form of the fair price 
of American option in Karatzas (1988, p. 50). By construction, 
$\tilde C_{B,A}$ does not allow arbitrage.
\enr

\enex

%In Example \ref{ex:toy}, the ``fair'' price is the maximum of {\it B}-prices
%over all probabilities. 

%\ber \label{r:manyEMM}
%When there are many equivalent martingale probabilities (EMP's),
%we can assume each buyer of the option believes in one of them. For 
%simplicity assume we have two EMP's, $P_1^*$ and $P_2^*. $
%Let us call the corresponding prices $C_1$ and $C_2$ with 
%$C_1 < C_2.$ If the market price of the option is $C_1,$
 %then the $P_2^*$-believer thinks there is an 
%arbitrage opportunity because of the expected 
%payoff at maturity under $P_2^*.$ How can you make all buyers believe there is
 %no arbitrage? Choose as price the $\max(C_1, C_2).$

%One can obtain the same result 
%directly by contradiction using the definition of 
%arbitrage. If there is arbitrage, we  buy at $t_0$ the option by paying C and 
%the payoff at maturity T is strictly larger than $Ce^{r(T-t_0)}$ with positive 
%probability and equal 
%to $Ce^{r(T-t_0)}$ otherwise.  
%The inequality also holds for the expected payoff under 
%anyone of 
%the EMP's and discounting both sides at $t_0$ 
%we 
%obtain on the left side  the "fair" price under this EMP that is strictly larger
%than C. This cannot happen if C is the maximum of all 
%the  prices obtained under the EMP's. 
%\enr

Conditions are given below for prices which follow Geometric model that 
guarantee the process $\{S_t/E_PS_t, \ t>0\}$ is a martingale under $P.$ In 
the sequel it is seen that these
conditions hold when the stock price process is modeled by a Geometric Brownian
motion or Hyperbolic Levy motion.

\bel \label{l:PMartingale}
For the stock price process $\{S_t, \ t>0\}$ on the probability
space $(\Omega, {\cal F}, P)$ assume that
\begin{equation}
%S_t=s_0e^{\mu t + V_t}, \ t>0, \mbox{ with } Ee^{V_t}=M^t, \ M>0,
S_t=s_0e^{\mu t + V_t}, \ t\ge 0, 
\label{eq:PMartingale}
\end{equation}
with $\mu \in R, \ V_0=0$  
and $\{V_t, \ t>0\}$ having stationary and independent increments.
Then, $E_Pe^{V_t}=M^t, \ M>0,$ and
the mean-adjusted prices $\{S_t/E_PS_t, \ t>0\}$  are 
martingale under $P.$
\enl

\section{$\tilde C_{B,t_0}(P)$ for Geometric Brownian motion}

\quad It is seen that $\tilde C_{B,t_0}(P)$ is the {\it B-S-M}-price $C;$ there
is no need to determine $P^*.$
Additional justification is initially provided
for the use of
discounting factor $A^{-1}(t_0,T)=s_{t_0}/E_PS_T.$
Recall that in the {\it B-S-M} assumptions, the stock price process $\{S_t, \ t>0\}$
on $(\Omega, {\cal F}, {P})$
satisfies the stochastic differential
equation
\begin{equation}
dS_t=\mu S_t dt+ \sigma S_t dW_t, \ t>0,
\label{eq:bsmsde}
\end{equation}
with $\{W_t, \ t >0 \} $ one-dimensional standard Brownian motion and
$\{{\cal F}_t, \ t >0 \}$ the natural filtration.

In Musiela and Rutkowski (1997, p. 110-111)
it is shown that for $t<T,$
\begin{equation}
E_P(S_T|{\cal F}_{t})=E_P(S_T|S_{t})=S_{t} e^{\mu (T-t)}.
\label{eq;projectTtot}
\end{equation}
The coefficient $e^{\mu (T-t)}$ describes the evolution
of the price process from $t$ to $T$ and by taking expected values
in (\ref{eq;projectTtot}) it follows that
the discounting factor from $T$ to $t$ is
\begin{equation}
e^{-\mu (T-t)}=\frac{E_PS_t}{E_PS_{T}}.
\label{eq:discTtot}
\end{equation}

When the starting time is $t_0,$ equation (\ref{eq:bsmsde}) has analytic solution
\begin{equation}
S_t=s_{t_0}exp\{ (\mu-\frac{\sigma^2}{2})(t-t_0)+\sigma W_{t-t_0}\}, \ t>t_0;
\label{eq:gbm}
\end{equation}
$s_{t_0}$ is the share's price at
$t_0.$

From (\ref{eq:discTtot}) and (\ref{eq:gbm}),
to discount the share's price from $T$ at $t_0$
we  use
$A^{-1}(t_0,T)=\frac{E_PS_{t_0}}{E_PS_T}=\frac{s_{t_0}}{E_PS_T}.$

\bef
For densities $f$ and $g$ on the real line, their Hellinger distance $H(f,g)$
is defined by
\begin{equation}
\label{eq:Hellinger}
H^2(f,g)=\int_R \{ \sqrt{f(x)}-\sqrt{g(x)}\}^2 dx.
\end{equation}
\enf

We then have the following proposition. 

\bep \label{c:bmbsm}
Under ({\cal A}) and the {\it B-S-M} assumptions, discounting
$S_T$ with $A^{-1}(t_0,T)$ and cash with $e^{-r(T-t_0)}$  and
defining $ F_0, \ f_0, \ f_1, \ F_1$ for $P$ as in (\ref{eq:h0}) and 
(\ref{eq:F11cdf}), it is shown that
the Bayes price  $\tilde C_{B,t_0}(P)$ 
of the European call option 
is the {\it B-S-M} price. The Hellinger distance $H(f_1,f_0)$ increases
with the volatility.
% at $t_0$ of the European option with strike price $X$ at maturity $T.$ 
 \enp 

\ber
When the price process is a Geometric Brownian motion under P,
from (\ref{eq:gbm}) for $t>u>t_0$ it holds
$$E_P[S_t/E_PS_t|{\cal F}_u]=e^{-.5\sigma^2(u-t_0)+\sigma W_u}
E_P[e^{-.5\sigma^2(t-u)+
\sigma W_{t-u}}]
=S_u/E_PS_u,$$
i.e. the mean-adjusted prices $\{S_t/E_PS_t, t>t_0\}$ are a martingale under
P. Alternatively, observe that
$E_Pe^{\sigma W_t}=(e^{.5\sigma^2})^t$ and Lemma \ref{l:PMartingale}
holds with $M=e^{.5\sigma^2}.$
\enr

%\bec 
%Under the {\it B-S-M} assumptions and Corollary \ref{c:bmbsm}, consider a game
%G that has loss $s_{t_0}-C$ with probability $1-R_B$ and profit 
%$C+Xe^{-r(T-t_0)}$
%with probability $R_B$ (the Bayes risk). From (\ref{eq:Bayesprice}), the 
%{\it B-S-M}-price $\tilde C_{B,t_0}$ makes $G$ a 
%``fair''-game and it is determined by calculating $R_B.$
%\enc

%\section{Bayes price for other models}
\section{{\it B}-prices for other models}

\quad It is widely known that the distribution of the logarithm of price returns  deviates from normality and
the constant volatility assumption is often violated. Thus, researchers 
use  also
normal mixtures,
distributions with heavier than normal tails, hyperbolic returns etc. 
{\it B}-price of the call option can be informative in these situations.

Assume that $t_0=0$ and that the stock price follows the model
\begin{equation}
S_t=s_{0}e^{\mu t+ X_t},  EX_t=0, \ t>0. 
\label{eq:model1}
\end{equation}
%$\mu=\mu_T, \ \sigma=\sigma_T$ and 

It is seen below that when $X_t$ in (\ref{eq:model1}) is a normal mixture,
mean-adjusted prices are not martingale under $P.$ However, for small
$t$-values they ``nearly'' are and in a 2-normal mixture
example it is observed that  
{\it B-S-M} price,  obtained assuming $X_t$ is Brownian motion
with the same mean and variance, is often larger than 
%the associated
{\it B}-price  and the mixture of {\it B-S-M}  prices obtained for 
each normal in the mixture.
%smaller than the {\it B-S-M} price  obtained assuming $X_t$ is Brownian motion.
When $X_t$ is a hyperbolic L\'{e}vy  motion
%the market is not complete and 
the martingale probability is not unique  but mean-adjusted prices are martingale under $P$
and a ``fair"
{\it B}-price is obtained.

\subsection{The normal mixture model}

\quad When $X_t$ in (\ref{eq:model1}) follows a normal mixture,
\begin{equation}
{\cal L}(X_t|P)=\sum_{i=1}^m p_i{\cal N}(0,a_i^2t),
\label{eq:mixtureN}
\end{equation}
$0<p_i<1, \ a_i>0, \ i=1,\ldots,m, \ \sum_{i=1}^m p_i=1.$
Then,
$$E_PS_t=s_{0}e^{\mu t}E_Pe^{X_t}=s_0e^{\mu t} \sum_{i=1}^m p_i 
e^{\frac{a_i^2t}{2}},$$
$$\frac{S_t}{E_PS_t}=e^{X_t - \ln  \sum_{i=1}^m p_i e^{\frac{a_i^2t}{2}} }, \mbox{ and}$$
$$f_0={\cal L}(\ln \frac{S_T}{E_PS_T}|P)=\sum_{i=1}^m p_i {\cal N}(-\ln G_T, a_i ^2T)$$
with
\begin{equation}
G_T=\sum_{i=1}^m p_i e^{\frac{a_i^2T}{2}}, \ t_0=0.
\label{eq:lnmean}
\end{equation}
Recall that $f_1(y)=e^{y}f_0(y)$  and observe that
$$E_{f_0}e^Y=\sum_{i=1}^mp_i e^{-ln G_T + \frac{a_i^2T}{2}}=1.$$
Then,
$$f_1(y)=e^{y}f_0(y)=\sum_{i=1}^m p_i\frac{1}{a_i \sqrt{2\pi T}} e^{-\frac{y^2+(\ln G_T)^2+2y(\ln G_T - a_i^2T)}{2a_i^2T}}
$$
$$=\sum_{i=1}^m p_i e^{\frac{-2a_i^2 T \ln G_T +a_i^4T^2}{2a_i^2T}} {\cal N}(-\ln G_T + a_i^2T, a_i^2T)
=\sum_{i=1}^m \frac{p_i e^{ \frac{a_i^2 T}{2}} }  {G_T}  {\cal N}(-\ln G_T + a_i^2T, a_i^2T).$$
From (\ref{eq:Bayes2}),
%(\ref{Bayespricelocscale}), 
{\it B}-price of the European call option is
\begin{equation}
\tilde C_{B,0}(P)=s_{0}\sum_{i=1}^m q_i\Phi(\frac{D-\ln G_T+a_i^2T}{a_i \sqrt{T}})
-Xe^{-rT}\sum_{i=1}^m p_i \Phi(\frac{D-\ln G_T}{a_i \sqrt{T}}),
\label{Bayespricemix}
\end{equation}
with $D=\ln(s_{0}/X)+rT, \  q_i=\frac{p_i e^{ \frac{a_i^2 T}{2}} }  {G_T}, 
\ i=1,\ldots, m$ and $G_T $ as in (\ref{eq:lnmean}).
%{\sum_{i=1}^m p_i e^{ \frac{a_i^2 \sigma^2}{2}}}.$
We observe that
 $\{S_t/E_PS_t, \ t>0\}$ are not  martingale for $m>1.$ 
However, for small $t\ (>u)$ they ``nearly'' are since 
$$E_P(S_t/E_PS_t|{\cal F}_u)=\frac{S_u}{E_PS_u}
\frac{ \sum_{i=1}^m p_i e^{a_i^2 u} \ 
\sum_{i=1}^m p_i e^{a_i^2 (t-u)} }
{\sum_{i=1}^m p_i e^{a_i^2t}}
\approx_{\mbox{small }t} \frac{S_u}{E_PS_u}.$$
In Example \ref{ex:calc1}, {\it B}-price  is computed for small $T$ and is compared with the
{\it B-S-M} price for Geometric Brownian motion with the 
same variance and the corresponding mixture of {\it B-S-M} prices obtained 
under the mixture model.
%Frequently the {\it B-S-M} price is larger than the $B$-price for the true model, which is the ``fair''  price for small 
%$T$ values, 
%thus indicating overpricing due to the wrong model.

\beex
\label{ex:calc1}
Assume that prices $\{S_t, t>0\}$ follow model (\ref{eq:model1}) and
$m=2$ in the mixture model (\ref{eq:mixtureN}), $p_1=p=1-p_2.$
We examine the effect of contaminations of the Geometric Brownian motion 
in the {\it B-S-M} price by comparing it with the {\it B}-price obtained under
the  mixture distribution for small $T$-values.
We use $i=.04, \ .08, \ t_0=0$ and small but also larger values
$T=.03,\ .05,\ .1,\ .15,\ .2,\ .5$ (in years),
$s_0=60 \ \$, \ X=70 \ \$, \ a_1=1, \
a_2=1.05, \ 1.2, \ 2, \ 4$  and the mixing coefficient $p$ takes values
$j/50, j=1,\ldots,50.$ The {\it B-S-M} price is 
obtained assuming prices 
follow
Geometric Brownian motion with variance $pa_1^2+(1-p)a_2^2.$ 
The results in Tables 1 and 2 indicate that the {\it B}-price 
is often smaller than the 
{\it B-S-M} price when $T$ is small; this does not hold when $T=.5.$
It has been observed, for example, that  {\it B-S-M} price is larger than {\it B}-price  
and the same mixture of {\it B-S-M} prices 
obtained for each model 
in the mixture; 
$t=0,\ T=.02, \ s=60, \ X=50, \ i=.08, \ K=50,  \ a_1=1, \ a_2=3.$ 
\enex
% \mbox{ } \\
\begin{center}
\begin{tabular}{|c|c|c|c|c|}\hline
\multicolumn{5}{|c|}
{\bf HOW OFTEN {\it B}-PRICE $<$ {\it B-S-M} PRICE, i=.04 }\\
\hline T (in years) & a2=1.05 & a2=1.2 & a2=2 &a2=4 \\
\hline .03& 0.01960784 & 0.01960784 & 0.9607843 & 0.4509804 \\
\hline .05& 0.9607843 & 0.9803922  & 0.9607843 &0.2352941  \\
\hline .1 & 0.9607843 & 0.9803922  & 0.7647059 & 0.01960784  \\
\hline .15& 0.9803922  &  0.9803922 & 0.4313725  & 0.01960784    \\
\hline .2 & 0.9607843  & 0.9607843 & 0.1568627 & 0   \\
\hline .5 & 0.01960784 & 0.01960784 & 0.01960784  & 0.01960784   \\
\hline
\end{tabular}\\
\mbox{ }\\
{\bf TABLE 1}
\end{center}

\begin{center}
\begin{tabular}{|c|c|c|c|c|}\hline
\multicolumn{5}{|c|}
{\bf HOW OFTEN {\it B}-PRICE $<$ {\it B-S-M} PRICE, i=.08 }\\
\hline T (in years) & a2=1.05 & a2=1.2 & a2=2 &a2=4 \\
\hline .03& 0.01960784  & 0.1764706  & 1 & 0.4705882 \\
\hline .05& 0.9803922 & 0.9803922  & 1 & 0.2745098 \\
\hline .1 & 0.9607843   & 0.9607843 & 0.8039216 & 0.01960784 \\
\hline .15& 0.9607843  &  0.9803922 & 0.4313725  & 0    \\
\hline .2 & 0.9803922  & 0.9607843 & 0.1960784  & 0   \\
\hline .5 & 0.01960784 & 0.03921569 & 0.01960784 & 0.01960784   \\
\hline
\end{tabular}\\
\mbox{ }\\
{\bf TABLE 2}
\end{center}

\subsection{The Hyperbolic L\'{e}vy motion model}

\quad Barndorff-Nielsen (1977) introduced the family of hyperbolic continuous distributions
with logarithmic  densities  hyperbolas. Barndorff-Nielsen and Halgreen (1977)
showed these are infinitely divisible. From empirical findings on stock returns, 
Eberlein and Keller (1995) (E \& K) considered 
the L\'{e}vy process $\{Z_t, \ t>0\},$ defined by the infinitely divisible
hyperbolic distribution that is symmetric and centered with density 
\begin{equation}
h(x;\zeta, \delta)=\frac{1}{2\delta K_1(\zeta)} e^{-\zeta \sqrt{1+(\frac{x}{\delta})^2} }, \ x \in R;
\label{eq:LMD}
\end{equation}
$K_1$ is the modified Bessel function of the third kind with index $1.$  
The process $\{Z_t, \ t>0\}$ has stationary, independent increments such that $Z_0=0,$ 
$Z_1$ has density $h(x;\zeta, \delta)$ and characteristic function $\phi(u;\zeta, \delta),$
and $Z_t$ has density
$$f_t(x;\zeta, \delta)=\frac{1}{\pi}\int _0^{\infty}cos(ux)\phi^t(u;\zeta,\delta)du.$$
E \& K called $\{Z_t, \ t>0\}$ hyperbolic L\'{e}vy motion 
and used (\ref{eq:model1}) with $\mu_t=0$ to model stock prices,
\begin{equation}
S_t=s_{0}e^{Z_{t}}, \ t>0.
\label{eq:levymodel}
\end{equation}
E \& K noticed that for model (\ref{eq:levymodel})
% is not complete and
 there is 
no unique martingale probability and obtained a price for the European 
call option under a martingale probability using the Esscher transform 
of the process. Recent Fourier transform valuation formulas for L\'{e}vy
and other models and securities can be found in
Eberlein, Glau and Papantoleon (2010).

%By stationarity and  independence of the increments, 
%the moment generating function
%of $Z_t$  (E\& K, p. 297, equation 30)
%$$E_Pe^{Z_t}=(E_Pe^{Z_1})^t=
%(\frac{\zeta}{K_1(\zeta)}\frac{K_1(\sqrt{\zeta^2-\delta^2})}
%{\sqrt{\zeta^2-\delta^2}})^t:=M^t, \ \delta<\zeta,$$
%$$E_PS_t=s_{0}E_Pe^{Z_{t}}=s_{0}M^t.$$
A ``fair'' $B$-price is now obtained under $P$ complementing prices obtained  using the Esscher 
transform (Gerber and Shiu, 1994).

%next Corollary follows.
\bep
\label{p:LevyMG}
a) The mean-adjusted prices $\{S_t/ES_t, \ t>0\}$ are martingale under $P.$\\
b) $B$-price is
$$\tilde C_{B,0}(P)=s_0\int_{\ln (X/s_0)-rT}^{\infty}e^x f_T(x+T\ln M)dx
-Xe^{-rT}\int_{\ln (X/s_0)-rT}^{\infty} f_T(x+T\ln M)dx.$$
%$$=\frac{s_0}{M^T}\int_{\ln(X/s_0)-T(r-\ln M)}^{\infty}e^yf_T(y)dy
%-Xe^{-rT} \int_{\ln(X/s_0)-T(r-\ln M)}^{\infty}f_T(y)dy.$$
%and therefore we can compute
%the {\it B}-price for this model.
\enp
%{\it a)} By stationarity and  independence of the increments,
%the moment generating function
%of $Z_t$  (E\& K, p. 297, equation 30)
%$$E_Pe^{Z_t}=(E_Pe^{Z_1})^t=
%(\frac{\zeta}{K_1(\zeta)}\frac{K_1(\sqrt{\zeta^2-\delta^2})}
%{\sqrt{\zeta^2-\delta^2}})^t:=M^t, \ \delta<\zeta,$$
%$$E_PS_t=s_{0}E_Pe^{Z_{t}}=s_{0}M^t.$$
%{\it b)} From Corollary \ref{eq:LevyMG}  compute  {\it B}-price  as 
%described in section 2.
%We then have
%$$\ln \frac{S_T}{E_PZ_T}=Z_T - T \ln M \sim_P f_t(x+T \ln M)=f_0(x)$$
%and since $t_0=0$ from (\ref{eq:Bayes2}) the {\it B}-price is
%$$\tilde C_{B,0}(P)=s_0\int_{\ln (X/s_0)-rT}^{\infty}e^x f_T(x+T\ln M)dx
%-Xe^{-rT}\int_{\ln (X/s_0)-rT}^{\infty} f_T(x+T\ln M)dx.$$\\

%Consider the process 
%$$\{S_t/ES_t, t>t_0\}.$$
%Then, for $u<t,$
%$$E(S_t/ES_t|{\cal F}_u)=\frac{e^{Z_u}}{M(1,1)^t}E(e^{Z_{t-u}})$$
%$$=\frac{e^{Z_u}}{M(1,1)^t}M(1,1)^{t-u}=\frac{e^{Z_u}}{M(1,1)^u}=
%\frac{S_u}{ES_u},$$
%and this implies that $\{S_t/ES_t, t>t_0\}$ is martingale.

\section{\bf Concluding Remarks}

\quad A  purely statistical interpretation of the price $C$ of the European 
call option has been provided from new formula 
(\ref{eq:THEMAINRESULT1}).
Advantages of {\it B}-prices include: \\ {\it a)} When  
mean-adjusted stock prices $\{S_t/E_PS_t, t_0 \le t \le T\}$ are martingale 
under $P, \ C$ can be obtained without prior determination of $P^*$ 
as in sections 4 and 5.2. \\
{\it b)} For small $T$-values, $\{S_t/E_PS_t\}$ is often nearly a martingale 
under $P$ and an approximation for $C$ can be obtained as in section 5.1.\\ 
{\it c)} In
(\ref{eq:THEMAINRESULT1}), one could model $R_B,$ estimate the unknown 
parameters using $C$ market-values and use the so-obtained estimate 
to derive other $C$-values.

\begin{center}
{\bf Acknowledgment}
\end{center}

Many thanks are due to my N.U.S. colleagues Professor Tiong Wee Lim  for stimulating conversations on option pricing and Professors Wei-Liem Loh and Sanjay Chaudhuri for their encouragement about this work. Many thanks are also due to Professor Tze Leung Lai 
for his encouragement about this work and to Professor Wolfgang H\"ardle for the careful reading of this manuscript and his suggestions to improve its readability.

\section*{Appendix}

{\bf Proof of Proposition \ref{p:P*fairisBayes}}  {\it a)} We start by 
proving the last equality in (\ref{eq:CRB}). Let 
%Use $R(d)$ to denote the  right  side  
%$R(d)$ 
%in  (\ref{eq:Bayes}),
%%%$$R(d)=\pi_1 F_1(\ln \frac{d}{ a(t_0,T)})
%%%+\pi_0 [1-F_0(\ln \frac{d}{ a(t_0,T)})].$$ 
\begin{equation}
\label{eq:BR}
R(d)=\pi_1 F_1^*(\ln \frac{d}{ES_T})
+\pi_0 [1-F_0^*(\ln \frac{d}{ES_T})]. 
\end{equation}
In the right side  of (\ref{eq:BR}), regions
 $(-\infty, \ln \frac{d}{ES_T}]$
and
 $(\ln \frac{d}{ES_T}, +\infty)$
are a partition of the real line  so they determine a decision function 
and 
$R(d)$ is Bayes risk for the estimation problem
of $F_1^*$ and $F_0^*$ with $0-1$ loss and prior probabilities
$\pi_1$ and $\pi_0$ respectively.
To minimize $R_d$ consider its first derivative, 
%%%$$R'(d)=\pi_1 F_1^*'(\ln \frac{d}{ a(t_0,T)})g(d)-\pi_0 F_0^*'(\ln \frac{d}{ a(t_0,T)})g(d)$$
$$R'(d)=\pi_1 F_1^{*'}(\ln \frac{d}{ES_T})g(d)-\pi_0 F_0^{*'}(\ln \frac{d}{ES_T})g(d)$$
%%%$$=g(d)\pi_0 f_0(\ln \frac{d}{a(t_0,T)})(\frac{\pi_1}{\pi_0} \frac{f_1(\ln \frac{d}{ a(t_0,T)})}
%%%{f_0(\ln \frac{d}{ a(t_0,T)})}-1)$$
$$=g(d)\pi_0 f_0^*(\ln \frac{d}{ES_T})(\frac{\pi_1}{\pi_0} \frac{f_1^*(\ln \frac{d}{ES_T})}
{f_0^*(\ln \frac{d}{ES_T})}-1)$$
\begin{equation}
%%%=g(d)\pi_0 f_0(\ln \frac{d}{ a(t_0,T)})(\frac{\pi_1}{\pi_0} \frac{d}{ka(t_0,T)}-1);
=g(d)\pi_0 f_0^*(\ln \frac{d}{ES_T})(\frac{\pi_1}{\pi_0} \frac{d}{ES_T}-1),
\label{eq:Bayesroot}
\end{equation}
where (\ref{eq:Bayesroot}) follows from (\ref{eq:F11cdf}); $g(d)$ 
is a term due to the first derivative.

Thus, from  (\ref{eq:F11cdf}) and (\ref{eq:Bayesroot})
%\begin{equation}
$$
%%%R'(d_B)=0 \Longleftrightarrow \pi_1 f_1(\ln \frac{d_B}{a(t_0,T)})=\pi_0 f_0(\ln \frac{d_B}{a(t_0,T)})
R'(d_B)=0 \Longleftrightarrow \pi_1 f_1^*(\ln \frac{d_B}{ES_T})=\pi_0 f_0^*(\ln \frac{d_B}{ES_T})
$$
\begin{equation}
%%% \Longleftrightarrow d_B=\frac{\pi_0 k a(t_0,T)}{\pi_1}=kX\frac{e^{-r(T-t_0)}}{s_{t_0}a^{-1}(t_0,T)}.
\Longleftrightarrow d_B=\frac{\pi_0  ES_T}{\pi_1}=X\frac{e^{-r(T-t_0)}}{s_{t_0}/ES_T}.
\label{eq:dBvalue}
\end{equation}
It also holds 
%%%$$R''(d_B)=g(d_B)\pi_1 f_0(\ln \frac{d_B}{ a(t_0,T)})\frac{1}{ka(t_0,T)}>0$$
$$R''(d_B)=g(d_B)\pi_1 f_0^*(\ln \frac{d_B}{ES_T})\frac{1}{ES_T}>0$$
and the minimum Bayes risk
\begin{equation}
\label{eq:BR1}
R_B=R(d_B)=
%R(\frac{Xe^{-r(T-t_0)}k_0 a(t_0,T)}{s_{t_0}})=
%%%\pi_1 F_1(\ln  \frac{Xe^{-r(T-t_0)}k}{s_{t_0}})
%%%+\pi_0 [1-F_0(\ln  \frac{Xe^{-r(T-t_0)}k}{s_{t_0}})]$$
\pi_1 F_1^*(\ln  \frac{Xe^{-r(T-t_0)}}{s_{t_0}})
+\pi_0 [1-F_0^*(\ln  \frac{Xe^{-r(T-t_0)}}{s_{t_0}})].
\end{equation}
To prove the first equality in (\ref{eq:CRB}),
note that since interest discounted stock prices are martingale under $P^*,$
\begin{equation}
e^{-r(T-t_0)}ES_T=s_{t_0} \rightarrow ES_T=s_{t_0}e^{r(T-t_0)}.
\label{eq:thekey}
\end{equation}
Use (\ref{eq:thekey}) to express the ``fair'' price $C$ of the option 
(in (\ref{eq:FAIRPRICESETUPA}) ) using
$F_1^*$ and $F_0^*,$ 
$$C=e^{-r(T-t_0)}ES_TI(S_T>X)-Xe^{-r(T-t_0)}P^*(S_T>X)$$
\begin{equation}
=s_{t_0} E e^{\ln \frac{S_T}{ES_T}}I(\ln \frac{S_T}{ES_T}>
\ln \frac{X}{s_{t_0}e^{r(T-t_0)}})-Xe^{-r(T-t_0)}P^*(\ln \frac{S_T}{ES_T}
>\ln \frac{X}{s_{t_0}e^{r(T-t_0)}})
\label{eq:intermcalc}
\end{equation}
$$=s_{t_0} [1-F_1^*(ln \frac{X}{s_{t_0}}-r(T-t_0))]
-Xe^{-r(T-t_0)} [1-F_0^*(ln \frac{X}{s_{t_0}}-r(T-t_0))],$$
and then
$$s_{t_0}-C=s_{t_0}F_1^*(ln \frac{X}{s_{t_0}}-r(T-t_0))+
Xe^{-r(T-t_0)}[1-F_0^*(ln \frac{X}{s_{t_0}}-r(T-t_0))] 
\le s_{t_0}+Xe^{-r(T-t_0)},$$ 
%with  $s_{t_0}+Xe^{-r(T-t_0)}$ an upper bound for the assets of the 
%writer's call who buys the share at $t_0$ and has liability $s_{t_0}-C,$
or
\begin{equation}
\label{eq:BR2}
\frac{s_{t_0}-C}{s_{t_0}+Xe^{-r(T-t_0)}}=
\pi_1F_1^*(ln \frac{X}{s_{t_0}}-r(T-t_0))+\pi_0
[1-F_0^*(ln \frac{X}{s_{t_0}}-r(T-t_0))].
\end{equation}
The result follows from (\ref{eq:BR1}) and (\ref{eq:BR2}).\\
{\it b)} Follows from part {\it a)} and (\ref{eq:BR1}).
%%% when $k=1.$
$\hspace{3ex} \hfill \Box$

{\bf Proof of Corollary  \ref{c:locscaprice}}
Follows from (\ref{eq:Bayes2}) and the assumption
$G_i(x)=1-G_i(-x), \ x \in R, \  i=0, 1.$
$\hspace{3ex} \hfill \Box$   

{\bf Proof of Corollary \ref{c:simplegame}} Follows from 
Proposition \ref{p:P*fairisBayes} rearranging (\ref{eq:Bayesprice}) to
obtain 
$$R_B(C+Xe^{-r(T-t_0)})-(s_{t_0}-C)(1-R_B)=0. \hspace{3ex} \hfill \Box$$

{\bf Proof of Lemma \ref{l:PMartingale}} From model (\ref{eq:PMartingale})
it holds
$$E_Pe^{V_t}=(E_Pe^{V_1})^t=M^t \mbox{ and }$$
$$E_PS_t=s_0e^{\mu t}E_Pe^{V_t}=s_0e^{\mu t}M^t,$$
and therefore, for $t>u,$ from stationarity and independence of $\{V_t\}$
increments
$$E_P(\frac{S_t}{E_PS_t}|{\cal F}_u)=\frac{e^{V_u}}{M^t}{E_Pe^{V_{t-u}}}
=\frac{e^{V_u}}{M^u}=\frac{S_u}{E_PS_u}. \hspace{3ex} \hfill \Box $$

%{\bf Proof of Proposition \ref{p.minimax}} Consider portfolio
%$$V_d=e^{-r(T-t_0)}(S_T-X)I(S_T>d).$$
%For any $Q$ in ${\cal P},$
%$$E_{P^*}e^{-r(T-t_0)}(S_T-X)I(S_T>d) \le E_Qe^{-r(T-t_0)}(S_T-X)I(S_T>d)$$
%$$= \frac{E_Q S_T e^{-r(T-t_0)}}{s_{t_0}} 
%s_{t_0}E_Q\frac{S_TI(S_T>d)}{E_QS_T}-Xe^{-r(T-t_0)}Q(S_T>d)$$
%$$=[\frac{E_Q S_T e^{-r(T-t_0)}}{s_{t_0}}-1]s_{t_0}E_Q\frac{S_TI(S_T>d)}{E_QS_T}
%+s_{t_0}E_Q\frac{S_TI(S_T>d)}{E_QS_T}-Xe^{-r(T-t_0)}Q(S_T>d)$$
%and taking the supremum over $d$
%$$C_{B,t_0}(P^*)\le E_Q S_T e^{-r(T-t_0)}-s_{t_0}+C_{B,t_0}(Q) \mbox { and}$$
%$$C_{B,t_0}(P^*) \le \inf_{Q \in {\cal P}-P^*} [E_Q S_T e^{-r(T-t_0)}-s_{t_0}
%+C_{B,t_0}(Q)].\hspace{3ex} \hfill \Box$$
%$$E_Qe^{-r(T-t_0)}(S_T-X)I(S_T>d)\ge E_{P^*}e^{-r(T-t_0)}(S_T-X)I(S_T>d)$$
%and therefore
%$$\frac{E_Q S_T e^{-r(T-t_0)}}{s_{t_0}} C_{B,t_0}(Q) 
%\ge C_{B,t_0}(P^*), \mbox{ and}$$
%$$inf_{Q \in {\cal P}}\frac{E_Q S_T e^{-r(T-t_0)}}{s_{t_0}}C_{B,t_0}(Q)
%=C_{B,t_0}(P^*). \hspace{3ex} \hfill \Box$$

{\bf Proof of Proposition \ref{c:bmbsm}} From Remark \ref{r:fromP*toP}, 
%under $P$ 
the value
of the writer's expected
cost at $T$ discounted at $t_0$ 
%with the suggested discounting under $P$ 
is
given by (\ref{eq:intermcalc}) with $P^*$ replaced by $P$ and
therefore (\ref{eq:Bayes2}) holds with
$F_0$ and $F_1$ instead of $F_0^*$ and $F_1^*.$
From (\ref{eq:gbm}) it follows that
$E_PS_T=s_{t_0}exp\{\mu (T-t_0)\},$ and
$${\cal L}(\ln \frac {S_T}{E_PS_T}|P)
={\cal N} (-\frac{\sigma^2}{2}(T-t_0), \sigma^2(T-t_0))=F_0;$$
${\cal N}(\theta, \tau^2)$ is used to denote a normal distribution with
mean $\theta$ and variance $\tau^2.$
% that $k=1$ and
From (\ref{eq:F11cdf}) it follows that
%with $\phi$ denoting standard normal density
%(\ref{eq:h1}) 
$$f_1(x)=e^x \frac{1}{\sigma \sqrt{T-t_0}}\phi(\frac{x+.5\sigma^2(T-t_0)}{\sigma
^2(T-t_0)})=
 \frac{1}{\sigma \sqrt{T-t_0}}\phi(\frac{x-. 5\sigma^2(T-t_0)}{\sigma^2(T-t_0)})
$$
$\mbox{i.e. } F_1= {\cal N}(\frac{\sigma^2(T-t_0)}{2}, \sigma^2(T-t_0));
 \ \phi$ denotes standard normal density.
%and $k_0=1.$

From (\ref{Bayespricelocscale}), with $G_1=G_0={\cal N}(0,1), 
\ \theta_1=\frac{
\sigma^2(T-t_0)}{2}, \   \theta_0=-\frac{\sigma^2(T-t_0)}{2},$ it follows that
%\[
 \begin{equation}
\tilde C_{B, t_0}(P)=s_{t_0}\Phi(d_1)-Xe^{-r(T-{t_0})}
\Phi(d_2),
\label{eq:bsmprice}
\end{equation}
%\]
%\begin{equation}
\[
d_j=\frac{\ln (s_{t_0}/X)+r(T-t_0)+ (-1)^{1+j}\  \frac{\sigma^2}{2}(T-t_0)}
{\sigma \sqrt{T-t_0}}=\frac{D+ (-1)^{1+j}\  \frac{\sigma^2}{2}(T-t_0)}
{\sigma \sqrt{T-t_0}}, \ j=1,2;
%\label{eq:d12}
%\end{equation}
\]
%$s_{t_0}$ is the share's observed price at $t_0,$
$\Phi$ is the cumulative distribution of a standard normal, $\sigma >0, \  D$ is
determined in (\ref{eq:Bayesbarrier}). 
It follows that 
$$H^2(f_1,f_0)=2(1-e^{-\frac{\sigma^2(T-t_0)}{8}}). \ \hspace{3ex} \hfill \Box$$

%{\bf Proof of Corollary  \ref{eq:LevyMG}}
{\bf Proof of Proposition  \ref{p:LevyMG}}
{\it a)} By stationarity and  independence of the increments,
the moment generating function
of $Z_t$  (E\& K, p. 297, equation 30)
$$E_Pe^{Z_t}=(E_Pe^{Z_1})^t=
(\frac{\zeta}{K_1(\zeta)}\frac{K_1(\sqrt{\zeta^2-\delta^2})}
{\sqrt{\zeta^2-\delta^2}})^t:=M^t, \ \delta<\zeta,$$
$$E_PS_t=s_{0}E_Pe^{Z_{t}}=s_{0}M^t.$$
The result follows from Lemma
%\ref{l:LevyMG}.
\ref{l:PMartingale}.

{\it b)} Compute  {\it B}-price  as 
described in section 2.
We then have
$$\ln \frac{S_T}{E_PZ_T}=Z_T - T \ln M \sim_P f_T(x+T \ln M)=f_0(x)$$
and since $t_0=0$ from (\ref{eq:Bayes2}) the {\it B}-price is
$$\tilde C_{B,0}(P)=s_0\int_{\ln (X/s_0)-rT}^{\infty}e^x f_T(x+T\ln M)dx
-Xe^{-rT}\int_{\ln (X/s_0)-rT}^{\infty} f_T(x+T\ln M)dx.$$
%$$=\frac{s_0}{M^T}\int_{\ln(X/s_0)-T(r-\ln M)}^{\infty}e^yf_T(y)dy
%-Xe^{-rT} \int_{\ln(X/s_0)-T(r-\ln M)}^{\infty}f_T(y)dy.$$
$\hspace{3ex} \hfill \Box$

%%%\begin{figure}
%\resizebox{4in}{6in}{
%\resizebox{5in}{6in}{
%\resizebox{6in}{7in}{
%%%\resizebox{7in}{8in}{
%%%\includegraphics{BAYESBOUNDNORMALFig1.ps}}
%%%\end{figure}

%TAKEN AWAY FOR THE ARCHIVES \begin{figure}
%\resizebox{6in}{7in}{
%\includegraphics{BAYESBSMMIXBSMFig2.ps}}
%\end{figure}

\end{document}